\documentclass[11pt]{article}
\usepackage{graphicx}
\usepackage{amsmath,amsthm,amssymb}
\begin{document}
\title{\textbf{\textsf{Interacting Holographic Viscous Dark Energy Model }}}
\author{ Mubasher Jamil\footnote{mjamil@camp.edu.pk}\ \ \ and M. Umar
Farooq\footnote{m\_ufarooq@yahoo.com}
\\ \\
\small Center for Advanced Mathematics and Physics, National University of Sciences and Technology,\\
\small E\&ME Campus, Peshawar Road, Rawalpindi, 46000, Pakistan \\
} \maketitle
\begin{abstract}
In this manuscript, we present a generalization of the interacting
holographic dark energy model using the viscous generalized
Chaplygin gas. We also study the model by considering a dynamical
Newton's constant $G$. Then we reconstruct the potential and the dynamics of the scalar field which describe the viscous Chaplygin cosmology.
\end{abstract}

\newpage
\large

\section{Introduction}

Astrophysical observations suggest that the observable universe is undergoing in a
transition from the earlier deceleration phase (matter dominated
era) to the acceleration phase (dark energy dominated era)
\cite{perl}. The empirical results based on the data are
$\Omega_m\approx0.3$ while $\Omega_\Lambda\approx0.7$. The simplest
explanation to dark energy phenomenon is the cosmological constant
$\Lambda$ which satisfies the equation of state (EoS) $p=-\rho$ or
$\omega=-1$ ($p=\omega\rho$) where $p$ is pressure and $\rho$ is the energy density. If the universe is dominated by $\Lambda$ than it will
maintain constant energy density and will dominate matter rapidly at
some instant in the evolution of the universe. Although the cosmological constant
offers a solution to the dark energy problem but with several
drawbacks like fine tunning and cosmic coincidence problem. The
former one requires fine tunning of the the energy density of dark energy to match the theoretical and observational values. The later
problem arises since it is unlikely that the current transition
period coincides with the current time. In other words, why the energy
densities of matter and dark energy are so much comparable at the
current epoch.

One of the promising resolutions to the dark energy problem is the
model based on the dark energy-dark matter (DE-DM) interaction and
it presents a possible resolution to the above problems \cite{jamil}. The interaction is
assumed to be negligible at high redshifts while it is large at
lower redshifts, thus it is motivating to make observations to detect
the said interaction \cite{bertolami}. A possible way to alleviate
the coincidence problem is to suppose that there is an interaction
between matter and dark energy. The cosmic coincidence can then be
alleviated by appropriate choice of the form of the interaction
between matter and dark energy leading to a nearly constant ratio
$r_m \equiv \rho_m/\rho_{de}$ during the present epoch or giving
rise to attractor of the cosmic evolution at late time. In the
current model, it can be rephrased as why the interaction rate is of
the order of the Hubble rate at present epoch. Recently some tracker
solutions are obtained for the DE-DM model which show that once the
attractor for the system is reached, the ratio between the
corresponding energy densities remains constant afterwards, thereby
solving the coincidence problem \cite{franca}. It is also recently
studied the dark energy decay into matter at the Hubble rate which
is a good fit with the observational data supporting an accelerating
universe \cite{berger}.

In this paper, we offer a connection between the holographic viscous
dark energy and the interacting dark energy. The former model is
also an alternative proposal to the problem of cosmic accelerated
expansion. In the second last section, we extend our study by
considering a dynamical Newton's constant $G$. Finally we conclude
our paper.

\section{The interacting model}

We start by assuming the background to be spatially homogeneous and
isotropic Friedmann-Robertson-Walker (FRW) spacetime, specified by
the line element:
\begin{equation}
ds^2=-dt^2+a^2(t)\left[
\frac{dr^2}{1-kr^2}+r^2(d\theta^2+\sin^2\theta d\phi^2) \right],
\end{equation}
where $a(t)$ is the scale factor and $k$ is the curvature parameter. For the sake of generality, we shall assume $k$ to be different from zero. The corresponding Einstein field equation is given by
\begin{equation}
H^2+\frac{k}{a^2}=\frac{1}{3M_p^2}\left[\rho_{de}+\rho_m\right].
\end{equation}
Here $M_p^2=(8\pi G)^{-1}$ is the reduced Planck mass. We assume
matter and dark energy interacting each other, then the energy
conservation equations read
\begin{eqnarray}
\dot{\rho}_{de}+3H(\rho_{de}+p_{de})&=&-Q,\\
\dot{\rho}_m+3H\rho_m&=&Q.
\end{eqnarray}
Here overdot represents differentiation with respect to cosmic time $t$. In explicit form, we have $p_{de}=\omega_{de}\rho_{de}$ and $p_m=0$
(or $\omega_m=0$). Note that the subscripts $de$ and $m$ refer to dark energy and matter respectively. Due to energy transfer, local energy
conservation will not hold but for the whole interacting system,
thus interaction leads to a modification of the standard $\Lambda$CDM
model. This interaction is naturally expected if the two species
exist in dominant quantities. It is generally assumed that
baryons don't interact dark energy and dark matter can. Since both dark energy and dark matter are largely unknown,
therefore the precise expression for the interaction would be
largely hypothetical. Here
$Q(\alpha_m\rho_m,\alpha_{de}\rho_{de})\simeq\alpha_m\rho_m+\alpha_{de}\rho_{de}$
is the interaction term which is a function of densities and two
coupling parameters corresponding the interacting components
\cite{sadjadi}. It determines the rate of change of energy in the
unit comoving volume. The direction of transfer of energy depends on
the sign of $Q$ i.e. positive $Q$ represents energy transfer from
dark matter to dark energy and vice versa for negative $Q$. Since
more parameters make the model to be less and less predictive, so we
shall use $\alpha_m=\alpha_{de}=b^2$ \cite{karwan}. In \cite{ma}, it
is suggested that interaction term should be proportional to the
number densities of the interacting medium to get a physically
interesting interaction term. The interacting model also best fits
the data of luminosity distance of supernovae of type Ia and with
the WMAP observations of cosmic microwave background \cite{szy}.
These observations constrain the interacting parameter $b^2<10^{-2}$
at 3$\sigma$ level. Moreover, $Q=3Hb^2(\rho_m+\rho_{de})$ is the
explicit form of interaction will be used here onwards. Here $3H$ is
attached for dimensional consistency.

Here we define the effective equations of state for dark energy and
matter as \cite{kim}
\begin{equation}
\omega_{de}^{eff}=\omega_{de}+\frac{\Gamma}{3H},\ \
\omega_m^{eff}=-\frac{1}{r_m}\frac{\Gamma}{3H}.
\end{equation}
Here $\Gamma=Q/\rho_{de}$ is the decay rate. Making use of Eq. (5)
in (3) and (4) yields
\begin{eqnarray}
\dot{\rho}_{de}+3H(1+\omega_{de}^{eff})\rho_{de}&=&0,\\
\dot{\rho}_m+3H(1+\omega_m^{eff})\rho_m&=&0.
\end{eqnarray}
The dimensionless density parameters corresponding to matter, dark
energy and curvature are
\begin{equation}
\Omega_m=\frac{\rho_m}{\rho_{cr}}=\frac{\rho_m}{3H^2M_p^2},\ \
\Omega_{de}=\frac{\rho_{de}}{\rho_{cr}}=\frac{\rho_{de}}{3H^2M_p^2},\
\ \Omega_k=\frac{k}{(aH)^2}.
\end{equation}
Here $\rho_{cr}\equiv3M_p^2H^2$ is the critical density.
Observations indicate that universe is spatially flat but after the
inclusion of higher order corrections, spatial curvature enters the
luminosity distance of SN Ia supernova. In this connection, it is
demonstrated that the reconstruction of the EoS of dark energy can
lead to gross errors \cite{clarkson}.

In an isotropic and homogeneous FRW universe, the dissipative
effects arise due to the presence of bulk viscosity $\xi$ in cosmic
fluids. The theory of bulk viscosity was initially investigated by
Eckart \cite{eckart} and later on pursued by Landau and Lifshitz
\cite{landau}. Dark energy with bulk viscosity has a peculiar
property to cause accelerated expansion of phantom type in the late
evolution of the universe \cite{brevik}. It can also alleviate
several cosmological puzzles like age problem \cite{ricci},
coincidence problem \cite{problem} and phantom crossing \cite{ivor}.
A viscous dark energy EoS is specified by
\begin{equation}
p_{eff}=p_{de}+\Pi.
\end{equation}
Here $\Pi=-\xi(\rho_{de})u^\mu_{;\mu}$ is the viscous pressure and
$u^\mu$ is the four-velocity vector. We require $\xi>0$ to get
positive entropy production in conformity with second law of
thermodynamics \cite{sun}. In FRW model, it takes the form
$\Pi=-3H\xi$ \cite{pun}, so that
\begin{equation}
p_{eff}=\frac{\chi}{\rho_{de}^\alpha}-3H\xi(\rho_{de}).
\end{equation}
The first term on the right hand side is called the generalized Chaplygin gas with $0<\alpha\leq1$. It reduces to the Chaplygin gas if $\alpha=1$ and converts to polytropic case if $\alpha<0$. In general, $\xi(\rho_{de})=\nu\rho_{de}^s$, $\nu\geq0$, where $s$ and $\nu$ are constant parameters. In particular, for the case, $s=1/2$ i.e. $\xi(\rho_{de})=\nu\rho_{de}^{1/2}$, yields a power-law expansion for the scale factor  \cite{barrow}. Moreover, if we demand to have the occurrence of a big rip in the future cosmic time then we have the following constraint on the parameter $\nu$: $\sqrt{3}\nu>\beta$, where $\beta\equiv1+\omega_{de}$, leading the scale factor to blow up in a finite time \cite{cataldo}. We assume the parametric form
$\xi(\rho_{de})=\nu\rho_{de}^{1/2}$. Hence Eq. (10) becomes
\begin{equation}
p_{eff}=\frac{\chi}{\rho_{de}^\alpha}-3\nu H\rho_{de}^{1/2}.
\end{equation}
Use of Eq. (11) in the energy conservation equation,
$\dot\rho_{de}+3H(\rho_{de}+p_{eff})=0$, yields
\begin{equation}
\rho_{de}=\left[
\frac{Da^{-3(1+\alpha)(1-\nu\gamma)}-\chi}{1-\nu\gamma}
\right]^{\frac{1}{1+\alpha}}.
\end{equation}
Here $D$ is a constant of integration,
$\gamma=M_p^{-1}\sqrt{1-r_m}$, where
$r_m\equiv\rho_m/\rho_{de}=\Omega_m/\Omega_{de}$. In the absence of
interaction, $r_m$ will be decreasing with time and increasing in
the case of interaction. The effective EoS of dark energy is given
by \cite{setare3}
\begin{equation}
\omega_{de}^{eff}=-\frac{1}{3}-\frac{2\sqrt{\Omega_{de}-c^2\Omega_k}}{3c}.
\end{equation}
It is now clear that the current accelerated expansion is not the
first time in the expansion history of the universe rather it was
earlier preceded by cosmic inflation. The latter was supposedly driven
by a dynamically evolving scalar field commonly called inflaton
\cite{zhang}. It has motivated to develop scalar field models
dealing with minimally coupled scalar field. Hence it can be
anticipated that the current accelerated expansion is driven by a
similar dynamical scalar field $\phi$ with potential $V(\phi)$,
related to the energy density and the pressure of viscous dark
energy as
\begin{equation}
\rho_\phi=\frac{1}{2}\dot{\phi}^2+V(\phi)=\left[
\frac{Da^{-3(1+\alpha)(1-\nu\gamma)}-\chi}{1-\nu\gamma}
\right]^{\frac{1}{1+\alpha}}.
\end{equation}
\begin{eqnarray}
p_\phi=\frac{1}{2}\dot{\phi}^2-V(\phi)=\frac{\chi}{\left[
\frac{Da^{-3(1+\alpha)(1-\nu\gamma)}-\chi}{1-\nu\gamma}
\right]^{\frac{\alpha}{1+\alpha}}}-3\nu H\sqrt{\left[
\frac{Da^{-3(1+\alpha)(1-\nu\gamma)}-\chi}{1-\nu\gamma}
\right]^{\frac{1}{1+\alpha}}}.
\end{eqnarray}
Until recently, the interaction of dark energy based on the
holographic principle has been introduced to explain the coincidence
problem and phantom crossing scenario (the transition quintessence
to the phantom phase or the phantom non-phantom transition)
\cite{zimdahl}. The principle is based on the idea that all the
information contained inside a spatial volume can also be obtained
from the information present on its surface (see \cite{susskind} for
comprehensive review). The principle has emerged from the quantum
gravity of black holes. We shall follow the formulation of Cohen et
al \cite{cohen} who proposed that in quantum field theory a short
distance (UV) cut-off is related to a long distance (IR) cut-off due
to the limit set by forming a black hole. In other words, if the
quantum zero-point energy density $\rho_{de}$ is relevant to a UV
cut-off, the total energy of the whole system with size L should not
exceed the mass of a black hole of the same size, thus we have
$\rho_{de}L^3\leq M_p^2L$ \cite{zhang1}. After saturating this
inequality with the largest IR cut-off, we obtain the energy density
of the holographic dark energy is represented by
\begin{equation}
\rho_{de}=3c^2M_p^2L^{-2},
\end{equation}
Here $c$ is a small positive constant of order unity. From the
information of supernovae SN Ia and cosmic microwave background
radiation, it is deduced that the holographic parameter has the
constraint $c=0.81^{+0.23}_{-0.16}$ \cite{zhang2}. Also $L$ is the
infrared cut-off which can be taken as
\begin{equation}
L=ar(t).
\end{equation}
In the cosmological context, the largest IR cut-off can be taken as
the size of the Hubble horizon, $L=H^{-1}$. Using the FRW metric,
one can obtain \cite{setare}
\begin{equation}
L=a(t)\frac{\text{sinn}[\sqrt{|k|}R_h/a(t)]}{\sqrt{|k|}},
\end{equation}
where $R_h$ is the size of the future event horizon defined as
\begin{equation}
R_h=a(t)\int\limits_t^\infty\frac{dt^\prime}{a(t^\prime)}=a(t)\int\limits_0^{r_1}\frac{dr}{\sqrt{1-kr^2}}.
\end{equation}
The last integral has the explicit form as
\begin{equation}\int\limits_0^{r_1}\frac{dr}{\sqrt{1-kr^2}}=\frac{1}{\sqrt{|k|}}\text{sinn}^{-1}(\sqrt{|k|}r_1)=
\begin{cases} \text{sin}^{-1}(r_1) , & \, \,k=1,\\
             r_1, & \, \,  k=0,\\
             \text{sinh}^{-1}(r_1), & \, \,k=-1,\\
\end{cases}\end{equation}
The EoS parameter gives
\begin{eqnarray}
\omega_{de}^{eff}&=&\frac{p_{eff}}{\rho_{de}}+\frac{\Gamma}{3H}\\
&=&\frac{\chi}{\rho_{de}^{1+\alpha}}-3\nu
H\rho_{de}^{-1/2}+b^2\frac{(1+\Omega_k)}{\Omega_{de}},
\end{eqnarray}
or we can write
\begin{equation}
\chi=\rho_{de}^{1+\alpha}\left[\omega_{de}^{eff}+3\nu
H\rho_{de}^{-1/2}-b^2\frac{(1+\Omega_k)}{\Omega_{de}}\right].
\end{equation}
Inserting (13) in (23), we obtain
\begin{equation}
\chi=(3c^2M_p^2L^{-2})^{1+\alpha}\left[
-\frac{1}{3}-\frac{2\sqrt{\Omega_{de}-c^2\Omega_k}}{3c}+3H\nu(3c^2M_p^2L^{-2})^{-1/2}
-b^2\frac{(1+\Omega_k)}{\Omega_{de}}\right].
\end{equation}
Using (24) in (12), we get
\begin{equation}
D=a^{3(1-\nu\gamma)(1+\alpha)}(3c^2M_p^2L^{-2})^{1+\alpha}\left[
\frac{2}{3}-\nu\gamma-\frac{2\sqrt{\Omega_{de}-c^2\Omega_k}}{3c}+3H\nu(3c^2M_p^2L^{-2})^{-1/2}
-b^2\frac{(1+\Omega_k)}{\Omega_{de}}\right].
\end{equation}
From Eqs. (14) and (15), the kinetic term becomes
\begin{eqnarray}
\dot{\phi}^2&=&\rho_\phi+p_\phi,\nonumber\\
&=&\frac{1}{\rho_{de}^\alpha}[\rho_{de}^{1+\alpha}+\chi-3\nu H\rho_{de}^{\alpha+\frac{1}{2}}],\nonumber\\
&=&(3c^2M_p^2L^{-2})\Big[
\frac{2}{3}-\frac{2\sqrt{\Omega_{de}-c^2\Omega_k}}{3c}
-b^2\frac{(1+\Omega_k)}{\Omega_{de}}\Big].
\end{eqnarray}
Also, the potential term becomes
\begin{eqnarray}
2V(\phi)&=&(\rho_{de}-p_{eff}),\nonumber\\
&=&\frac{1}{\rho_{de}^\alpha}(\rho_{de}^{1+\alpha}-\chi+3\nu H\rho_{de}^{\alpha+1/2}).\nonumber\\
&=&(3c^2M_p^2L^{-2})\Big[
\frac{4}{3}+\frac{2\sqrt{\Omega_{de}-c^2\Omega_k}}{3c}
+b^2\frac{(1+\Omega_k)}{\Omega_{de}}\Big].
\end{eqnarray}
From (26), we have
\begin{equation}
\dot{\phi}=HM_p\left[3\Omega_{de}\Big\{
\frac{2}{3}-\frac{2\sqrt{\Omega_{de}-c^2\Omega_k}}{3c}
-b^2\frac{(1+\Omega_k)}{\Omega_{de}} \Big\}\right]^{1/2}.
\end{equation}
 Using the relation with
$x=\ln a$ \cite{setare2}, we have
\begin{equation}
\dot{\phi}=\phi^\prime H,
\end{equation}
where prime denotes differentiation with respect to the e-folding time parameter $x$, we obtain
\begin{equation}
\phi^\prime=M_p\left[3\Omega_{de}\left\{\frac{2}{3}-\frac{2\sqrt{\Omega_{de}-c^2\Omega_k}}{3c}
-b^2\frac{(1+\Omega_k)}{\Omega_{de}}\right\}\right]^{1/2}.
\end{equation}
On integration, we get
\begin{eqnarray}
\phi(a)-\phi(a_o)&=&\int\limits_o^{\ln
a}M_p\Big[3\Omega_{de}\Big\{\frac{2}{3}-\frac{2\sqrt{\Omega_{de}-c^2\Omega_k}}{3c}
-b^2\frac{(1+\Omega_k)}{\Omega_{de}} \Big\}\Big]^{1/2}.
\end{eqnarray}
It is interesting to note that the above expressions for the kinetic
and potential for the viscous dark energy are the ones as they were
for the non-viscous case.
\subsection{Interacting holographic viscous dark energy with variable Newton's constant $G$}

Now we perform the above analysis with considering variable $G$ i.e.
$G=G(a)$ and $G=G(t)$. There is some evidence of a variable $G$
through numerous astrophysical observations \cite{ray}. Models with
variable $G$ can fix some of the hardest problems in cosmology like
the age problem, cosmic coincidence problem and determination of the precise value of
the Hubble parameter \cite{vis}.

Differentiating Eq. (16), we obtain
\begin{equation}
\dot\rho_{de}=-\rho_{de}\Big(\frac{\dot G}{G}+2\frac{\dot L}{L}
\Big).
\end{equation}
Moreover, using the definitions $\Omega_{de}=\rho_{de}/\rho_{cr}$ and
$\rho_{cr}=3M_p^2H^2$, we can write
\begin{equation}
L=\frac{c}{H\sqrt{\Omega_{de}}}.
\end{equation}
Using Eq. (33) in (32), we obtain
\begin{equation}
\dot \rho_{de}=-\rho_{de}H\Big[
2-\frac{\sqrt{\Omega_{de}}}{c}\text{cosn}\Big(
\frac{\sqrt{|k|}R_h}{a} \Big) +\frac{G^\prime}{G}\Big].
\end{equation}
Substituting backwards Eq. (34) in the energy conservation equation
(6), we obtain
\begin{equation}
\omega_{de}^{eff}=-\Big[
\frac{1}{3}+\frac{2\sqrt{\Omega_{de}-c^2\Omega_k}}{3c}
\Big]+\frac{G^\prime}{3G}.
\end{equation}
Making use of (35) in (23), we get
\begin{equation}
\chi=(3c^2M_p^2L^{-2})^{1+\alpha}\left[
-\frac{1}{3}-\frac{2\sqrt{\Omega_{de}-c^2\Omega_k}}{3c}+3H\nu(3c^2M_p^2L^{-2})^{-1/2}+\frac{G^\prime}{3G}
-b^2\frac{(1+\Omega_k)}{\Omega_{de}}\right].
\end{equation}
Substituting (36) in (12), we get
\begin{equation}
D=a^{3(1-\nu\gamma)(1+\alpha)}(3c^2M_p^2L^{-2})^{1+\alpha}\left[
\frac{2}{3}-\nu\gamma-\frac{2\sqrt{\Omega_{de}-c^2\Omega_k}}{3c}+3H\nu(3c^2M_p^2L^{-2})^{-1/2}+\frac{G^\prime}{3G}
-b^2\frac{(1+\Omega_k)}{\Omega_{de}}\right].
\end{equation}
Similarly, the kinetic and potential terms modify to
\begin{equation}
\dot\phi^2=(3c^2M_p^2L^{-2})\Big[
\frac{2}{3}-\frac{2\sqrt{\Omega_{de}-c^2\Omega_k}}{3c}+\frac{G^\prime}{3G}
-b^2\frac{(1+\Omega_k)}{\Omega_{de}}\Big],
\end{equation}
\begin{equation}
2V(\phi)=(3c^2M_p^2L^{-2})\Big[
\frac{4}{3}+\frac{2\sqrt{\Omega_{de}-c^2\Omega_k}}{3c}+\frac{G^\prime}{3G}
+b^2\frac{(1+\Omega_k)}{\Omega_{de}}\Big].
\end{equation}
Thus we have reconstructed the kinetic and potential terms for the viscous HDE which involve variation in $G$.
\newpage
\section{Conclusion}
Holographic dark energy model presents a dynamical view of the dark
energy which is consistent with the astrophysical observations.
Different values of the holographic parameter $c$ correspond to
different values of the dark energy parameter $\omega_{de}$
\cite{jamil1}. Thus it gives a nice connection between the two
models. In this paper, we have constructed a correspondence between
holographic dark energy and interacting dark energy. This formalism
is made using the viscous generalized Chaplygin gas. This EoS
belongs to a general class of inhomogeneous EoS as suggested in
\cite{nojiri1}. Within the different candidates of dark energy, the Chaplygin gas has emerged as a possible unification of dark matter and dark energy, since its cosmological evolution is similar to an initial dust like matter and a cosmological constant for late times. We have found that the reconstruction of the kinetic and potential terms of the HDE are independent of the viscosity parameters. It implies that if the dark energy is of the holographic type then it will be non-viscous and non-dissipative. The viscosity effects at the cosmic scale, if any, will remain negligible in the evolution of holographic dark energy. Finally, we have constructed a
similar correspondence by considering a variable $G$. It shows that the variable gravitational constant will modify the evolution of the scalar field while viscosity effects remain negligible. 

\subsubsection*{Acknowledgment}
We would like to thank the referees for their useful criticism on this work.

\end{document}